\documentclass[aps,prb,twocolumn,superscriptaddress]{revtex4}
\usepackage{amssymb}
\usepackage{graphicx}
\usepackage{epsfig}
\usepackage{float}
\usepackage{amsmath}

\begin{document}

\title{The origin of a$_{1g}$ and e$_g$' orderings in Na$_x$CoO$_2$}
\author{D. Pillay}
\affiliation{Code 6393, Naval Research Laboratory, Washington, D.C. 20375}
\author{M.D. Johannes}
\affiliation{Code 6393, Naval Research Laboratory, Washington, D.C. 20375}
\author{I.I. Mazin}
\affiliation{Code 6393, Naval Research Laboratory, Washington, D.C. 20375}
\author{O.K. Andersen}
\affiliation{Max-Plank-Insitut f\"{u}r Fesk\"{o}rperforschung, Stuttgart, Germany}

\begin{abstract}
It has often been suggested that correlation effects suppress the small $
e_{g}^{\prime }$ Fermi surface pockets of Na$_{x}$CoO$_{2}$ that are
predicted by LDA, but absent in ARPES measurements. It appears that within
the dynamical mean field theory (DMFT) the ARPES can be reproduced only if
the on-site energy of the $e_{g}^{\prime }$ complex is lower than that of
the $a_{1g}$ complex at the one-electron level, \textit{prior} to the
addition of local correlation effects. Current estimates regarding the order
of the two orbital complexes range from -200 meV to 315 meV in therms of the
energy difference. In this work, we perform density functional theory
calculations of this one-electron splitting $\Delta $ = $\epsilon
_{a_{1g}}-\epsilon_{e_{g}^{\prime }}$ for the full two-layer compound, Na$
_{2x}$Co$_{2}$O$_{4}$, accounting for the effects of Na ordering,
interplanar interactions and octahedral distortion. We find that $\epsilon
_{a_{1g}}-\epsilon_{e_{g}^{\prime }}$
is negative for all Na fillings and that this is primarily due to the
strongly positive Coulomb field created by Na$^{+}$ ions in the intercalant
plane. This field disproportionately affects the $a_{1g}$ orbital which
protrudes farther upward from the Co plane than the $e_{g}^{\prime }$
orbitals. We discuss also the secondary effects of octahedral compression
and multi-orbital filling on the value of $\Delta $ as a function of Na
content. Our results indicate that if the $e_{g}^{\prime }$ pockets are
indeed suppressed that can only be due to nonlocal correlation effects
beyond the standard DMFT.
\end{abstract}

\maketitle

Since shortly after the discovery of superconductivity in hydrated 1.4H$_{2}$ O$\cdot $Na$_{0.35}$CoO$_{2}$ \cite{takada}, there has been 
controversy surrounding the discrepancy between the calculated and observed Fermi surfaces of the parent compound, Na$_{x}$CoO$_{2}$ 
($x=1/3)$. Density functional (DFT) calculations show a large central FS, surrounded by six small elliptical hole pockets \cite{DJS00}, 
while angular-resolved photoemission (ARPES) measurements find the large pocket but do not see the smaller ones 
\cite{H-Y04,MZH+03,qian06,yang05}. The presence or absence of these small pockets drastically changes the response properties of the 
system, and a number of proposed superconductivity models depend on them \cite{MDJ+b,mochizuki,HIYN04,kuroki}. There are two basic ways 
to resolve the controversy: either the experiment is not representative of the bulk electronic states, or correlation effects beyond the 
DFT calculations qualitatively change the Fermiology. The latter seems rather plausible, given the narrow width of the Co $t_{2g}$ bands 
and the fact that DFT fails to reproduce the nonmagnetic ground state for $x\lesssim 0.7$. The effects of adding correlation have been 
recently addressed through DMFT calculations which account for both on-site Coulomb repulsion (Hubbard $U$) and local fluctuation effects 
(albeit not for long-wave-length ferromagnetic calculations, compatible with the observed at $x\gtrsim 0.5$ magnetic interactions in this 
system). Despite initial controversy \cite {ishida,georges,marianetti}, it has been firmly established that DMFT calculations only agree 
with ARPES if the centroids of the the $a_{1g}$ and $e_{g}^{\prime }$ Co $d-$bands are such that $\epsilon _{a_{1g}}$ is higher than 
$\epsilon _{e_{g}^{\prime }}$, which would yield a positive splitting, $ \Delta =$ $\epsilon _{a_{1g}}-\epsilon_{e_{g}^{\prime }}>0$. If, 
on the other hand $\epsilon_{a_{1g}}-\epsilon_{e_{g}^{\prime }} <0$ then the small hole pockets are preserved and even slightly enlarged. 
This result does not depend on the value of U or the particular Hamiltonian used in the calculation \cite{liebsch}. The sign of $ \Delta 
$ is therefore of great importance, but difficult or impossible to access experimentally. A tight binding fit to full potential density 
functional theory calculations \cite{MDJ+b} employing all 5 Co $d$ orbitals and 3 O $p$ orbitals in a double layer system found a 
strongly negative $ \Delta \sim $ -200 meV for $x$ =0.3. Other tight binding models fit to a one-layer system and employing a smaller 
basis set of Co $t_{2g}$ orbitals \cite{ishida,us} found $\Delta \sim $ -144 meV, and an alternate procedure for extracting $\Delta$ from the band 
structure yielded a value of $\sim$ 100 meV \cite{georges}. On the other hand, quantum chemical calculations 
of embedded clusters \cite{landron1, landron2} invariably find a positive $\Delta $ = 315 meV. There has so far been little understanding 
of what might cause such strong discrepancy between the values obtained through the two different methodologies.

In this work, we present DFT calculations of $\Delta $ that include both layers of the double-layer unit cell and real Na ions (no 
virtual crystal or other approximation) and we extract our values directly from the DFT calculation without employing a tight-binding 
fit. It is crucial to understand that $\Delta $ is controlled entirely by the electrostatic field at the Co site and Co-O overlap 
integrals, which are both accurately described by DFT. Note that, within DFT, the Fermi surface itself hardly depends on the sign of 
$\Delta $, as long as $|\Delta |$ is moderate, but the effect of the local correlation effects (as evaluated by DMFT) is drastically 
different. At every Na doping level and for every reasonable value of octahedral compression, we find that $\Delta $ is negative. We find 
that the main cause of the ordering between $a_{1g}$ and $e_{g}^{\prime }$ orbitals is the positive Na potential in the interstitial 
plane that drives the out-of-plane a$_{1g}$ orbitals downward in energy, a factor that has been overlooked in all previous studies \cite{note2}. 
Interestingly, we find that octahedral distortion as an isolated factor \textit{increases} $\Delta $ , a reversal of crystal field 
predictions for CoO$_{6}$ octahedra. This highlights the strong role of Na to re-establish a negative $\Delta $ and emphasizes that Co-O 
hybridization depends on interactions beyond the CoO$ _{6}$ octahedra.

\subsection{Calculational Methods}

All calculations have been performed using a plane wave density functional
theory code, the Vienna Ab-initio Simulation Package (VASP) \cite{vasp}. The
projected augmented wave method (PAW) \cite{PAW1} was used along with the generalized gradient approximation \cite{Perdew:1996} to the 
exchange
correlation potential. A $\sqrt{3}\times \sqrt{3}$ cell that
preserves the overall hexagonal symmetry was used to simulate the bulk.
Ground state geometries were fully optimized using a Monkhorst Pack k-point
mesh of 4$\times $4$\times $2. For the total energy calculation and density
of states (DOS) the Brilliouin zone integrations were performed on a 8$
\times $8$\times $4 mesh.

To calculate $\Delta $, we first consider the Co plane in a cubic coordiate
system where three Co atoms are located equidistant along the $x,y,$ and $z$
axes, \textit{i.e.} they form a plane perpendicular to the [111] cubic axis
which is also the $z$ axis in the hexagonal system. In this cubic system, it
is easy to explicitly write out the $a_{1g}$ and $e_{g}^{\prime }$ orbitals:

\begin{figure} \includegraphics[width = 0.95\linewidth]{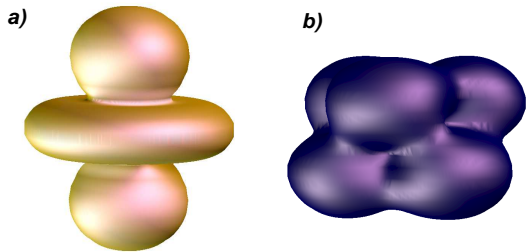} \caption{(color online) Charge density plots for a) the a$_{1g}$ and b) 
the e $_g$' complexes of NaCoO$_2$. The plots employ squared wavefunctions, produced using full DFT calculations, taken at the $\Gamma$ 
point where the a$_{1g}$ and e$_g$' states are pure.  The sum over both squared e$_g$' orbitals is shown in b).} \label{orb} \end{figure}

\begin{eqnarray}
a_{1g} &=&\frac{XY+YZ+XZ}{2}  \notag \\
e_{g}^{\prime } &=&XY-\frac{XZ-YZ}{2}  \label{transf} \\
e_{g}^{\prime } &=&\frac{\sqrt{3}(YZ-XZ)}{2}  \notag
\end{eqnarray}

Standard band structure codes project the density of states onto the three symmetry representations 
appropriate for a hexagonal symmetry, that is, $ N_{z^{2}}$ projects onto the $3z^{2}-1$ 
orbital, and $N_{x^{2}-y^{2},xy}$ and $N_{xz,yz},$ each give the total projection onto the two orbitals 
belonging to each respective representation. Since the projected density of states contains the 
square of the projected orbital, we square each term in Eq. \ref{transf}, keeping only the terms 
conforming to the correct symmetry. The resulting recipe is:

\begin{eqnarray}
N_{a_{1g}}(\epsilon ) &=&N_{z^{2}}(\epsilon ) \\
N_{e_{g}^{\prime }}(\epsilon ) &=&N_{x^{2}-y^{2},xy}(\epsilon
)+2N_{xz,yz}(\epsilon )
\end{eqnarray}

From these expressions, we calculate band centers of each complex. It is
important to note that the $e_{g}^{\prime }$ states share the same
irreducible representation as the unoccupied $e_{g}$ states and that there
is mixing between the two complexes \cite{landron2}. Therefore, without
specifically projecting out only the $e_{g}^{\prime }$ states as in our
methodology, erroneous values for $\Delta $ will be found. The onsite
parameters that result from this procedure are not identical to the onsite
parameters in the effective 3$\times 3$ Hamiltonian, so far used in DMFT
calculations because the latter contain not only the Co-based $a_{1g}$ and $
e_{g}^{\prime }$ states, but also the hybridized O states. The latter can be
obtained only by downfolding the calculated Bloch functions onto Wannier
functions, as has been done before \cite{ishida,zhou,georges,us}. Still, our
method can be used to estimate the changes in $\Delta $ brought about by
different factors, such as octahedral distortion, Na filling, or Na ordering. In the several
cases where we have compared our values of $\Delta $ to those calculated
from the Wannier functions of a three band model, we have found that the
latter are always \textit{more} negative than the values extracted directly
from DFT. It is worth noting that these Wannier functions are very long
range, having a sizeable weight not only on O and on nearest neighbor Co's,
but even at the third neighbor Co sites. This may have an effect on the
onsite energy ordering, although we believe it will be quantitative, not
qualitative. An additional effect of this extent is a strong reduction of the Hubbard $U$
as compared to typical values for local $d$-orbitals in
transition metal oxides.

\subsection{Results and Discussion}

\begin{figure}[tbp]
\includegraphics[angle = 270,width=0.95\linewidth]{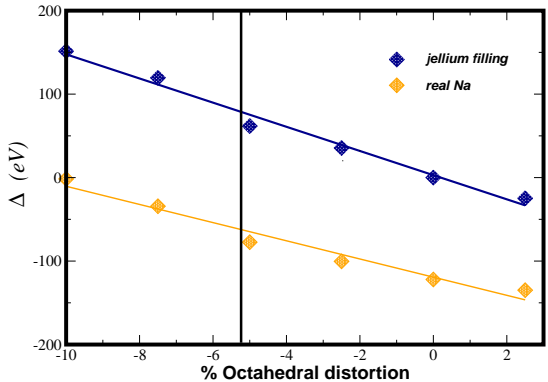}
\caption{(color online) $\Delta \equiv \epsilon_{a_{1g}} - \epsilon_{e_g'}$ as a function of 
octahedral distortion in
the jellium filled and real Na filled compounds. The vertical bar at $\sim$
5.2\% marks the position of the octahedral compression obtained through
relaxing the oxygen height in our calculations.}
\label{DC}
\end{figure}

We first address the effect of the octahedral compression (trigonal
distortion) that lowers the symmetry at the Co site and splits the $t_{2g}$
triplet into the $a_{{1g}}$ and $e_{g}^{\prime }$ complexes. Some previous
works have assumed that the distortion destabilizes the $a_{1g}$ state \cite
{korshunov,kroll,fcc04,wang}, while others assume oppositely that the $
e_{g}^{\prime }$ state is higher in energy \cite{zhou,WKSM03b}. We have
examined this question from two points of view. First, we have constructed a
tight-binding model of a CoO$_{6}$ octahedron, including the three Co $t_{2g}$
orbitals and three $p$ orbitals on each of the six O atoms. We have
diagonalized the resulting matrix for both compressive and expansive
distortions. Using the canonical scaling of the $t_{dp\sigma }$ and $
t_{dp\pi }$ parameters, namely $t_{dp\sigma }=\sqrt{3}t_{dp\pi}$, we find that the
compressive distortion, as seen in Na$_{x}$CoO$_{2}$ for all $x$, destabilizes
the $e_{g}^{\prime }$ state. However, if the ratio of the two parameters
were to be increased by approximately a factor of two, as has been seen in
some cases \cite{CDW}, the order of energies is reversed and the $a_{1g}$
state is destabilized. Second, we have performed a full DFT calculation of $\Delta $
as a function of the trigonal distortion. We use a fictitious compound, $
e^{-}$CoO$_{2}$, in which one electron per formula unit has been added to
the CoO$_{2}$ compound to achieve the band-insulating state, analogous to
NaCoO$_{2}$. The extra charge is compensated by a uniform positive
background. We use this technique, termed \textquotedblleft jellium
filling\textquotedblright , to eliminate the effects of the Na potential and of unequal band fillings,
which we discuss in later section.
As can be seen in Fig. \ref{DC}, octahedral compression always
produces a positive $\Delta $, \textit{i.e.} the $a_{1g}$ state is raised
above the $e_{g}^{\prime }$. This demonstrates that the scaling of $t
_{dp\sigma }$ and $t_{dp\pi }$ is outside the canonical range and points to
the importance of neighbors more distant than the surrounding O atoms.

By far the strongest factor determining the value of $\Delta $ is the
presence of ionized Na atoms in the intercalant plane. The positive Coulomb field
they produce serves to strongly shift the a$_{1g}$ orbital downward compared
to the $e_{g}^{\prime }$ orbital. As seen in Fig. \ref{orb}, the $a_{1g}$
orbital has a greater extent along the $z$ direction (c-axis of the crystal)
and therefore penetrates further into the Na plane. The resultant lowering
of the $a_{1g}$ states is sufficient to shift the sign of $\Delta $ from
positive to negative. Replacing the jellium filling discussed above with
real Na ions (again one per formula unit), we find a negative shift of
approximately 150 meV in $\Delta $ for all trigonal distortions, as seen in
Fig. \ref{DC}. Because quantum chemical calculations account only for the
filling effects of Na and neglect the electrostatic field produced by real
ions, this shift is not present and $\Delta $ is consequently found to be
strongly positive. Not surprisingly, when Na ions are located directly atop
a Co ion ($2b$ site), the effect is stronger than when they are located
equidistant between three Co ions ($2c$ site).

\begin{figure}[tbp]
\includegraphics[angle =270, width=0.95\linewidth]{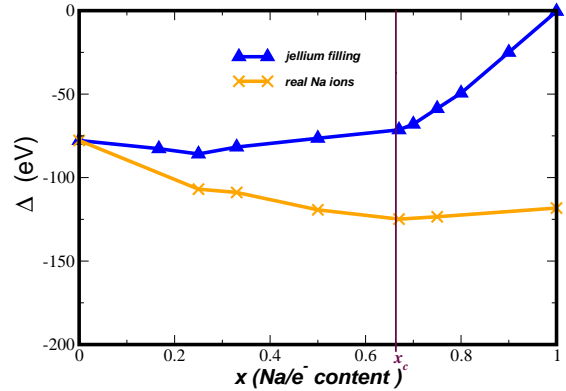}
\caption{(color online) $\Delta \equiv \epsilon_{a_{1g}} - \epsilon_{e_g'}$ as a function of 
electron filling for
jellium filled and real Na filled compounds. The critical doping $x_c$ where
the e$_g$' complex becomes completely filled is marked by a vertical bar.}
\label{multi}
\end{figure}

The final factor contributing to the size of $\Delta $ is the non-uniform filling of the two band complexes. For all levels of Na 
intercalation, the $ a_{1g}$ bands are less filled than the $e_{g}^{\prime }$ bands. At a critical filling of $x_{c}\sim $ 0.67, the 
$e_{g}^{\prime }$ complex becomes entirely filled, leaving holes only in the $a_{1g}$ bands. The increased on-site Coulomb repulsion due 
to greater filling drives the $e_{g}^{\prime }$ upward in energy compared to $a_{1g}$ bands. In the absence of real Na ions (hence no 
positive interstitial Coulomb field) and without octahedral distortion, the calculated $\Delta $ for CoO$_{2}$ is $-77$ meV, reflecting 
this multi-band effect. We have calculated $\Delta $ for Na$_{x}$CoO$_{2}$ and for $xe^{-}$CoO$_{2}$ (jellium filling) as a function of 
$x$, with all oxygen positions fixed to perfect octahedra. For the jellium filled compounds, $\Delta $ initially remains approximately 
constant and then increases dramatically at $x_{c}\sim $ 0.67 where the $e_{g}^{\prime }$ bands become entirely filled, as seen in Fig. 
\ref{multi}. The Na$_{x}$CoO$ _{2}$ curve, on the other hand, decreases initially and then flattens out once the $e_{g}^{\prime }$ 
complex is full. In the former case, electrons are added to both $e_{g}^{\prime }$ and $a_{1g}$ states at low $x$, pushing both complexes 
upward nearly equally. Once the $e_{g}^{\prime }$ holes are filled, all further electrons enter the $a_{1g}$ complex, pushing the 
$a_{1g}$ onsite energy up relative to $e_{g}^{\prime }$ and causing $ \Delta $ to increase. The situation is more complex when real Na 
ions are added. At low $x$, both $e_{g}^{\prime }$ and $a_{1g}$, electrons are added to both complexes and on-site repulsion raises both 
energies together, but the increasingly positive Coulomb field created by adding Na ions pushes the $a_{1g}$ states downward.  $\Delta $ 
therefore increases with $x$ as long as holes remain in both symmetry complexes. After $x_{c}$, where additional electrons enter only the 
$a_{1g}$ complex, the positive Coulomb field acts oppositely to the increased on-site repulsion. Apparently the magnitude of these two 
effects is approximately equal, as evidenced by the nearly constant $\Delta $ at $x>x_{c}$.

\begin{table}[ht]
\caption{$\Delta \equiv \epsilon_{a_{1g}} - \epsilon_{e_g'}$ values for full DFT calculations, 
including relaxation of
the lattice parameters and ionic positions. Values for Na placed at Na1 ($2b$
trigonal site) and at Na2 ($2c$ octahedral site) positions are given for
comparison.}
\label{table_del}\centering
\begin{tabular}{cccc}
\hline
$x$ & Na1 position & Na2 position &  \\ \hline\hline
$0$ & -80 meV & - &  \\ 
$1/3$ & -100 meV & -95 meV &  \\ 
$2/3$ & -114 meV & -101 meV &  \\ 
$1$ & -77 meV & -6 meV &  \\ \hline
\end{tabular}
\end{table}

The effect of octahedral distortion on $\Delta $ competes with the effect of
the positive Na$^{+}$ potential, and both are a function of the Na filling, $x
$, which separately affects the $a_{1g}$ and $e_{g}^{\prime }$ ordering.
Having studied each of these factors individually, we proceed to calculate $
\Delta $ as a function of $x$, accounting for all factors simultaneously.  This is accomplished
by relaxing the lattice parameters (both $a$ and $c$) and the ionic
positions at each Na filling and using supercell calculations to include real Na ions in the interstitial plane. Our results are
tabulated in Table \ref{table_del}.  Lechermann {\it et al} have derived $\Delta$ for $x$=0.5 with Na in the Na2 position,  by integrating the 
partial density of states and also by
constructing first principles Wannier functions, arriving at an estimate of -100 meV, in excellent agreement
with our Table 1. Their estimate for Delta at x=1 is also consistent with ours \cite{georges}. It is clear that $\Delta $ is 
a
non-monotonic function of $x$, owing to competition between the various effects associated with Na filling, but that it remains negative
for all Na content 0 $< x <$ 1.

\subsection{Conclusions}

The crystal field splitting $\Delta $ $\epsilon
_{a_{1g}}-\epsilon_{e_{g}^{\prime }}$ is known to be the parameter that
controls the presence or absence of the controversial $e_{g}^{\prime }$
pockets at the Fermi level in Na$_{x}$CoO$_{2}.$ Here we show that, within
the DFT framework which is highly reliable for calculating one-electron
parameters, $\Delta $ is always negative and strongly so for low $x$. We
have shown that this fact is mainly due to the electrostatic crystal field
of Na as well as demonstrating that the O ligand field (octahedral
distortion), in fact, leads to the opposite sign of $\Delta .$ The explicit
effect of ionized Na has been neglected or severely approximated in previous
calculations, leading erroneously to positive values of $\Delta $. This
result indicates that local correlation and/or fluctuation effects are not
responsible for suppressing the $e_{g}^{\prime }$ Fermi sheets. Either
longe-range fluctuations, outside the scope of DMFT, or strong deviations
from the bulk structure/stoichiometry such as surfaces, defects, or
vacancies, must therefore account for the non-observation of these pockets
in ARPES experiments.

\subsection{Acknowledgements} We would like to thank A. Liebsch for many helpful and interesting discussions and F. Lechermann for a critical 
reading of our manuscript and for sharing unpublished results.  Research at NRL is funded by the Office of Naval Research.  DP would like to 
acknowledge funding from a National Research Council Associateship.


\begin{thebibliography}{28}
\bibitem[{us()}]{us}
\bibinfo{note}{The same procedure used in Ref. \onlinecite{ishida} was used for
  $x$=0,0.1,and 0.5, yielding -151, -149, and -130meV}.
\bibitem[{not()}]{note2}
\bibinfo{note}{Similarly, it is the Coulomb potential of Mg or Ca that pushes
  down, respectively, the C $\pi$ band in MgB$_2$ and the free-electron band in
  CaC$_6$, which in both cases is instrumental in raising the critical
  temperature.}
\end{thebibliography}
\end{document}